\documentclass[twocolumn,showpacs,preprintnumbers,amsmath,amssymb,aps,prb,longbibliography]{revtex4-2}
\usepackage{graphicx}
\usepackage{dcolumn}
\usepackage{bm}
\usepackage{verbatim} 
\usepackage[usenames, dvipsnames]{color} 
\usepackage[dvipsnames]{xcolor}
\usepackage{ulem}
\usepackage{color}
\usepackage[colorlinks=false,pdfencoding=auto,psdextra]{hyperref}
\usepackage{multirow}
\usepackage{tabularx}
\usepackage{dsfont}

\definecolor{orange}{rgb}{1.0, 0.5, 0.0}

\begin{document}
\title{Effect of crystal symmetry of lead halide perovskites on the optical orientation of excitons}

\author{Nataliia~E.~Kopteva$^{1}$, Dmitri~R.~Yakovlev$^{1}$, Ey\"up~Yalcin$^{1}$, Ina V. Kalitukha$^{1}$, Ilya~A.~Akimov$^{1}$, Mikhail~O.~Nestoklon$^{1}$, Bekir~T\"uredi$^{2,3}$, Oleh~Hordiichuk$^{2,3}$, Dmitry~N.~Dirin$^{2,3}$, Maksym~V.~Kovalenko$^{2,3}$, and Manfred~Bayer$^{1}$} 
 
 \affiliation{$^{1}$Experimentelle Physik 2, Technische Universit\"at Dortmund, 44227 Dortmund, Germany}
\affiliation{$^{2}$Laboratory of Inorganic Chemistry, Department of Chemistry and Applied Biosciences,  ETH Z\"{u}rich, CH-8093 Z\"{u}rich, Switzerland}
\affiliation{$^{3}$Laboratory for Thin Films and Photovoltaics, Empa-Swiss Federal Laboratories for Materials Science and Technology, CH-8600 D\"{u}bendorf, Switzerland }

\date{\today}
\makeatletter
\newenvironment{mywidetext}{%
  \par\ignorespaces
  \setbox\widetext@top\vbox{%
   \hb@xt@\hsize{%
    \leaders\hrule\hfil
    \vrule\@height6\p@
   }%
  }%
  \setbox\widetext@bot\hb@xt@\hsize{%
    \vrule\@depth6\p@
    \leaders\hrule\hfil
  }%
  \onecolumngrid
  \vskip10\p@
  \dimen@\ht\widetext@top\advance\dimen@\dp\widetext@top
  \cleaders\box\widetext@top\vskip\dimen@
  \vskip6\p@
  \prep@math@patch
}{%
  \par
  \vskip6\p@
  \setbox\widetext@bot\vbox{%
   \hb@xt@\hsize{\hfil\box\widetext@bot}%
  }%
  \dimen@\ht\widetext@bot\advance\dimen@\dp\widetext@bot
  \vskip\dimen@
  \vskip8.5\p@
  \twocolumngrid\global\@ignoretrue
  \@endpetrue
}%
\makeatother

\begin{abstract} 
The great variety of lead halide perovskite semiconductors represents an outstanding platform for studying crystal symmetry effects on the spin-dependent properties. Access to them is granted through optical orientation of exciton and carrier spins by circularly polarized photons. Here, the exciton spin polarization is investigated at $1.6$\,K cryogenic temperature in four lead halide perovskite crystals with different symmetries: (almost) cubic in FA$_{0.9}$Cs$_{0.1}$PbI$_{2.8}$Br$_{0.2}$ and FAPbBr$_3$, and orthorhombic in MAPbI$_3$ and CsPbBr$_3$. Giant optical orientation of 85\% is found for the excitons in FA$_{0.9}$Cs$_{0.1}$PbI$_{2.8}$Br$_{0.2}$, MAPbI$_3$, and CsPbBr$_3$, while it amounts to 20\% in FAPbBr$_3$. For all studied crystals, the optical orientation is robust to detuning of the laser photon energy from the exciton resonance, remaining constant for high energy detunings up to 0.3~eV, above which it continuously decreases to zero for detunings exceeding 1~eV. No acceleration of the spin relaxation for excitons with large kinetic energy is found in the cubic and orthorhombic crystals. This evidences the absence of the Dyakonov-Perel spin relaxation mechanism, which is based on the Rashba-Dresselhaus splitting of spin states at finite $k$-vectors. This indicates that the spatial inversion symmetry is maintained in perovskite crystals, independent of the cubic or orthorhombic phase.
\end{abstract}

\maketitle

\subsection*{Keywords}
Lead halide perovskites, spintronics, excitons, optical spin orientation, time-resolved photoluminescence.

\section{Introduction}

The interest in studies of the optical properties of lead-halide perovskites is steadily growing due to their outstanding photovoltaic~\cite{jena2019} and optoelectronic properties~\cite{Vinattieri2021_book,Vardeny2022_book}. Their flexible chemical composition $A$Pb$X_3$, where the cation $A$ can be cesium (Cs$^+$), methylammonium (MA$^+$), formamidinium (FA$^+$), and the anion $X$ can be I$^-$, Br$^-$, Cl$^-$, provides tunability of the band gap from the infrared to the ultraviolet spectral range as well as temperature-dependent crystal symmetries~\cite{simanes2024}. The structural phase transition from cubic symmetry at high temperatures to orthorhombic one at cryogenic temperatures represents an excellent test bed for studying symmetry effects on spin-dependent properties which are important for applications in spintronics~\cite{Vardeny2022_book,wang2019,ning2020,kim2021}. In turn, the spin physics sheds light on details of the crystal symmetry, as spin polarization and spin dynamics are highly susceptible to it.

Many spin-related phenomena were explored in perovskite bulk, two-dimensional, and nanocrystal structures. The exciton and carrier spins can be optically oriented by circularly polarized light~\cite{Giovanni2015,XOO2024,COO2024}, which is exploited in experimental techniques measuring photoluminescence polarization, time-resolved Faraday/Kerr rotation~\cite{odenthal2017,belykh2019}, spin-flip Raman scattering~\cite{kirstein2022nc}, etc. The spin dynamics at cryogenic temperatures extend to nanoseconds for spin dephasing~\cite{kirstein2022AM} and to submilliseconds for longitudinal spin relaxation~\cite{Belykh2022}. The hyperfine interaction of excitons and carriers with the nuclear spin system is essential for the spin dynamics and relaxation~\cite{kirstein2022AM,Kirstein2023_DNSS,COO2024}, and limits the spin relaxation time in weak magnetic fields. As shown recently, the giant optical orientation of exciton spins in a FA$_{0.9}$Cs$_{0.1}$PbI$_{2.8}$Br$_{0.2}$ crystal reaches 85\%~\cite{XOO2024}, close to the ultimate limit of 100\%. This value is robust against detuning of the laser photon energy from the exciton resonance up to 0.3~eV, providing strong experimental evidence that the Dyakonov-Perel spin relaxation is absent and, therefore, the spin splitting in the conduction and valence bands caused by the spin-orbit interaction is zero in this particular crystal. 

In the lead halide perovskites, the spin-orbit coupling (SOC) is large due to the heavy lead ions, which is an essential prerequisite for the spin-dependent effects, emerging from the Rashba and Dresselhaus spin splittings of the electron and hole bands at finite wave vectors~\cite{Manchon2015}. The absence of spatial inversion symmetry is another key prerequisite for the SOC-induced effects. A comprehensive review of the Rashba and Dresselhaus effects in the perovskite semiconductors, including potential origins of inversion symmetry breaking, and experimental approaches for detecting these effects can be found in Refs.~\cite{Kepenekian2015,Kepenekian2017}. Theoretical predictions suggest strong Rashba-Dresselhaus splittings, but the mechanisms for symmetry breaking are still under debate. The lead halide perovskites inherently have inversion symmetry in their cubic crystal phase, and even when they transition to the tetragonal or orthorhombic phases, this inversion symmetry remains. The inversion symmetry can be broken at the crystal surface~\cite{Mosconi2017}, at boundaries between grains, and by application of external electric or strain field~\cite{Kepenekian2015,Leppert2016}. For single crystals, the possible existence of ferroelectric phases has been suggested as static mechanism for Rashba-Dresselhaus splittings~\cite{Kim2014,Yu2019}. For hybrid organic-inorganic perovskites, the orientation of the organic cations (MA$^+$, FA$^+$) and their dynamics may provide static or dynamic Rashba-Dresselhaus effects~\cite{Marronnier2018}, which are not expected for the fully-inorganic perovskites with Cs$^+$ cations. However, even for the fully inorganic CsPbI$_3$ crystals, vibrational instabilities were predicted to give rise to dynamic distortions of the lattice~\cite{McKechnie2018}. Given this unclear situation, unique experimental evidences are required to establish which mechanisms are relevant for which materials and for what experimental conditions.

The experiments available so far are insufficient for a comprehensive understanding of the bulk Rashba-Dresselhaus effects. A large Rashba splitting was observed in angle-resolved photoemission spectroscopy (ARPES) on a MAPbBr$_3$ single crystal~\cite{Niesner2016} and by two-photon absorption spectroscopy~\cite{Lafalce2022}. However, the ARPES technique is sensitive only to the symmetry breaking at the surface. The presence of the surface Rashba effect on a MAPbBr$_3$ single crystal was confirmed by the circular photogalvanic effect~\cite{Huang2021}. Later, the absence of the static Rashba-Dresselhaus splitting was evidenced by ARPES experiments on MAPbBr$_3$ and CsPbBr$_3$ crystals~\cite{Sajedi2020}, and by second harmonic generation experiments on MAPbBr$_3$ crystals~\cite{Frohna2018}.  In MAPbI$_3$ crystals, the dynamical Rashba effect caused by thermally induced structural fluctuations that break the inversion symmetry was evidenced by measuring the circular photogalvanic effect~\cite{Niesner2018}. The resulting spin splitting is the key basic aspect in the studied effects. Therefore, spin physics techniques, which allow one to study spin splittings and spin dynamics can greatly contribute to understanding this still unresolved problem.

In the present paper, we study the spin orientation and relaxation of excitons to clarify the effect of the spin-orbit coupling and identify a possible breaking of the inversion symmetry in various perovskite single crystals. We chose four lead halide perovskite crystals, FA$_{0.9}$Cs$_{0.1}$PbI$_{2.8}$Br$_{0.2}$, FAPbBr$_3$, MAPbI$_3$, and CsPbBr$_3$m that can be distinguished by: (i) different halogens (I and Br) determining the band gap energy, (ii) different cations (FA$^+$, MA$^+$, Cs$^+$), (iii) hybrid organic-inorganic and fully-inorganic perovskites, and (iv) different crystal symmetries at cryogenic temperatures ranging from almost cubic to orthorhombic. This crystal selection allows us to investigate the influence of these features on the optical orientation of exciton spins and on the exciton spin relaxation. In turn, we get information about the crystal symmetries, namely that the spatial inversion symmetry inherent to the cubic perovskite crystals is not broken in the orthorhombic crystals.

\section{Results and discussion}

Here, we examine four lead halide perovskite crystals with band gap energies ($E_\text{g}$) in the visible spectral range of $1.5 - 2.4$~eV and with different symmetries at cryogenic temperatures, namely FA$_{0.9}$Cs$_{0.1}$PbI$_{2.8}$Br$_{0.2}$, FAPbBr$_3$, MAPbI$_3$, and CsPbBr$_3$.

\subsection{Optical properties} 

\begin{figure*}[t]
\begin{center}
\includegraphics[width = 18cm]{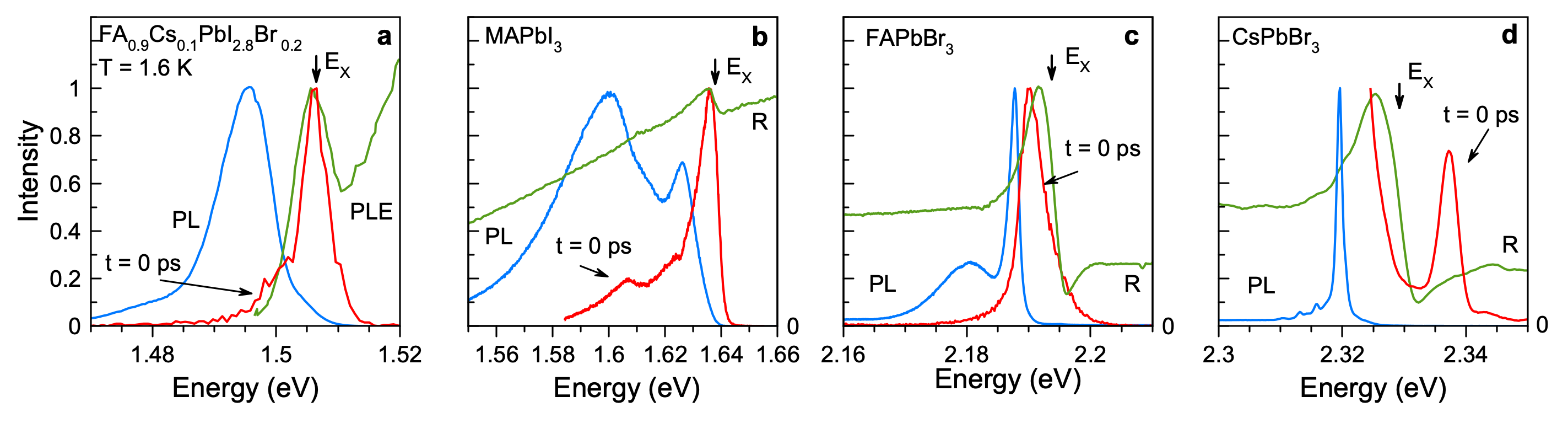}
\caption{\label{fig1} Optical properties of the studied lead halide perovskite crystals with various band gaps at 1.6~K temperature. The blue lines show time-integrated photoluminescence spectra measured for pulsed excitation at $E_\text{exc} = 3.06$~eV using the laser power density of $5$~mW/cm$^2$. The green lines show the PL excitation spectrum for FA$_{0.9}$Cs$_{0.1}$PbI$_{2.8}$Br$_{0.2}$, but reflectivity spectra for the other samples. $E_\text{X}$ marks the exciton resonance. PL spectra right after the excitation pulse $t = 0$~ps are depicted by the red lines. Laser excitation energy is $E_\text{exc} = 1.669$~eV for FA$_{0.9}$Cs$_{0.1}$PbI$_{2.8}$Br$_{0.2}$, $1.771$~eV for MAPbI$_3$, $2.296$~eV for FAPbBr$_3$, and $2.385$~eV for CsPbBr$_3$. }
\end{center}
\end{figure*}

The optical properties of the studied crystals at $T = 1.6$~K cryogenic temperature are collected in Figure~\ref{fig1}. The spectra of reflectivity (R) or photoluminescence excitation (the latter for FA$_{0.9}$Cs$_{0.1}$PbI$_{2.8}$Br$_{0.2}$) each show a pronounced exciton resonance, marked by an arrow and labelled with $E_\text{X}$. Time-integrated photoluminescence (PL) spectra shown by the blue lines demonstrate a Stokes shift of about $6-10$~meV from the energy of the exciton resonance. The emission is contributed by bound or localized excitons and by recombination of electrons and holes separated spatially~\cite{XOO2024,COO2024,deQuilettes2019_si,herz2017}. 

\begin{figure*}[htb]
\begin{center}
\includegraphics[width = 16cm]{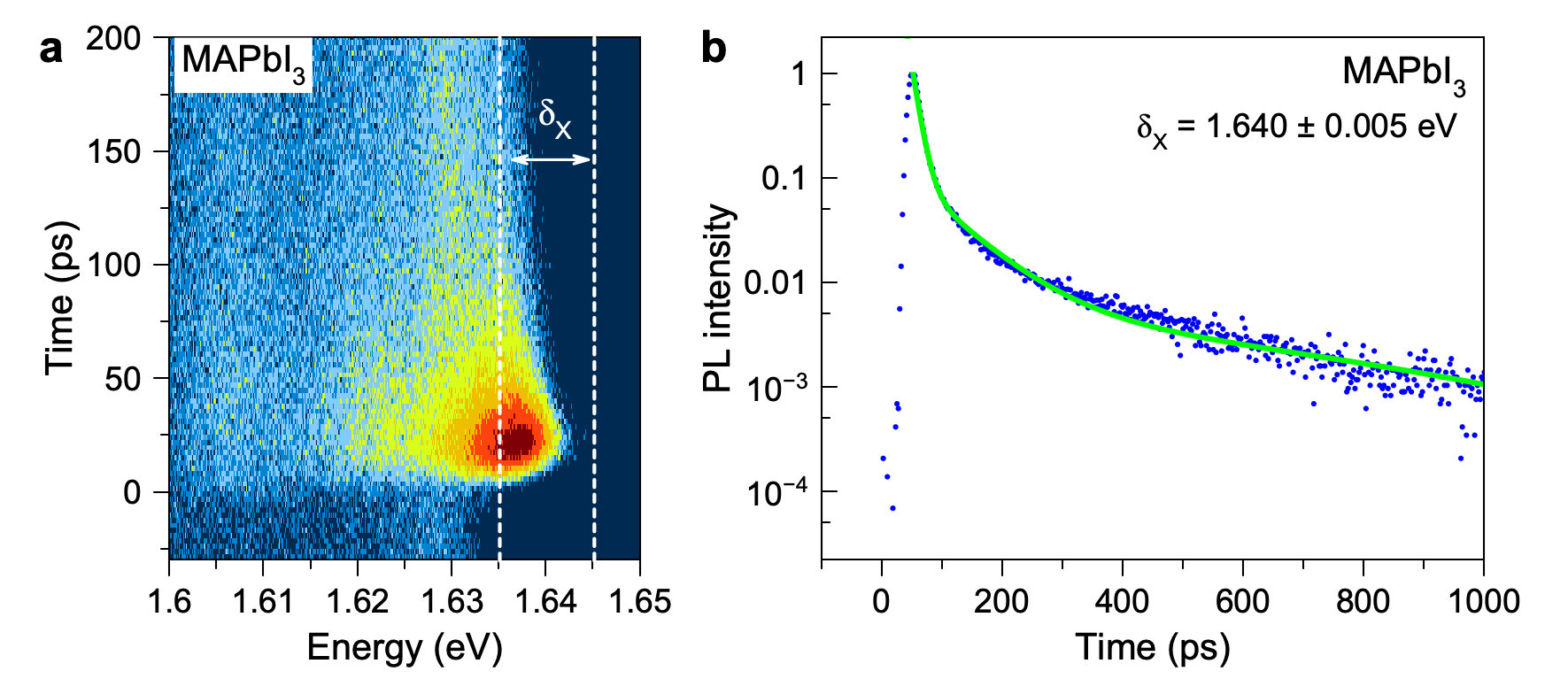}
\caption{\label{fig2} Recombination dynamics for the MAPbI$_{3}$ crystal at $T = 1.6$\,K. 
(a) Contour plot of time-resolved photoluminescence excited with 200\,fs laser pulses at the laser photon energy of 1.771~eV.  (b) Recombination dynamics integrated over the $\delta_\text{X} = 1.635-1.645$~eV spectral range around the exciton line maximum at $E_\text{X} = 1.636$\,eV. The green line is three-exponential fit with decay times: $\tau_{\rm R1} = 15$~ps, $\tau_{\rm R2} = 80$~ps, and $\tau_{\rm R3} = 810$~ps.}
\end{center}
\end{figure*}

We are interested in excitons with a short recombination time. Their emission can be isolated spectrally and temporally by measuring time-resolved PL using a streak camera, as shown in Figure~\ref{fig2}(a) for MAPbI$_{3}$.  Right after the excitation pulse action, the emission has its spectral maximum at $E_\text{X} = 1.636$\,eV, as demonstrated in Figure~\ref{fig1}(b) by the red line. The exciton dynamics spectrally integrated in the range of $1.635-1.645$~eV are shown in Figure~\ref{fig2}(b). The dynamics can be fitted with a three-exponential decay function. We assign the two short decay times of $\tau_{\text{R}1} = 15$~ps and $\tau_{\text{R}2} = 80$~ps to the exciton recombination. The long-lived component with $\tau_{\text{R}3} = 810$~ps is provided by recombination of spatially separated electrons and holes. It is a common feature for bulk perovskites that the recombination of electron-hole pairs and of excitons is spectrally overlapping~\cite{belykh2019,kirstein2022AM,kirstein2022nc,COO2024}. Ref.~\onlinecite{COO2024} shows how these contributions can be separated by employing the Hanle and polarization recovery effects in a magnetic field.

Spectrally, the exciton emission matches the resonance in the PL excitation spectrum for FA$_{0.9}$Cs$_{0.1}$PbI$_{2.8}$Br$_{0.2}$, and in the reflectivity spectra for MAPbI$_{3}$ and FAPbBr$_{3}$. Its time-integrated contribution is considerably weaker than the one of the Stokes shifted PL, so that in the time-integrated spectra (blue lines in Figure~\ref{fig1}), the exciton emission is not observed as separate line, but as high-energy shoulder. In the CsPbBr$_{3}$ crystal with a strong exciton-polariton effect~\cite{Yakovlev2024}, a separate emission line is seen arising from the upper polariton branch at 2.3375~eV, see the red line in Figure~\ref{fig1}(d). 

\subsection{Optical orientation of exciton spins}

The optical orientation of exciton spins induced by circularly polarized laser excitation is illustrated for the MAPbI$_3$ example. Note that the MAPbI$_3$ crystal has orthorhombic symmetry at $T=1.6$~K, which is well documented by its spin-dependent properties, including the strong anisotropy of the Land\`e $g$-factors of electron and hole~\cite{kirstein2022nc,kirstein2022mapi}. 

\begin{figure}[htb]
\begin{center}
\includegraphics[width = 8cm]{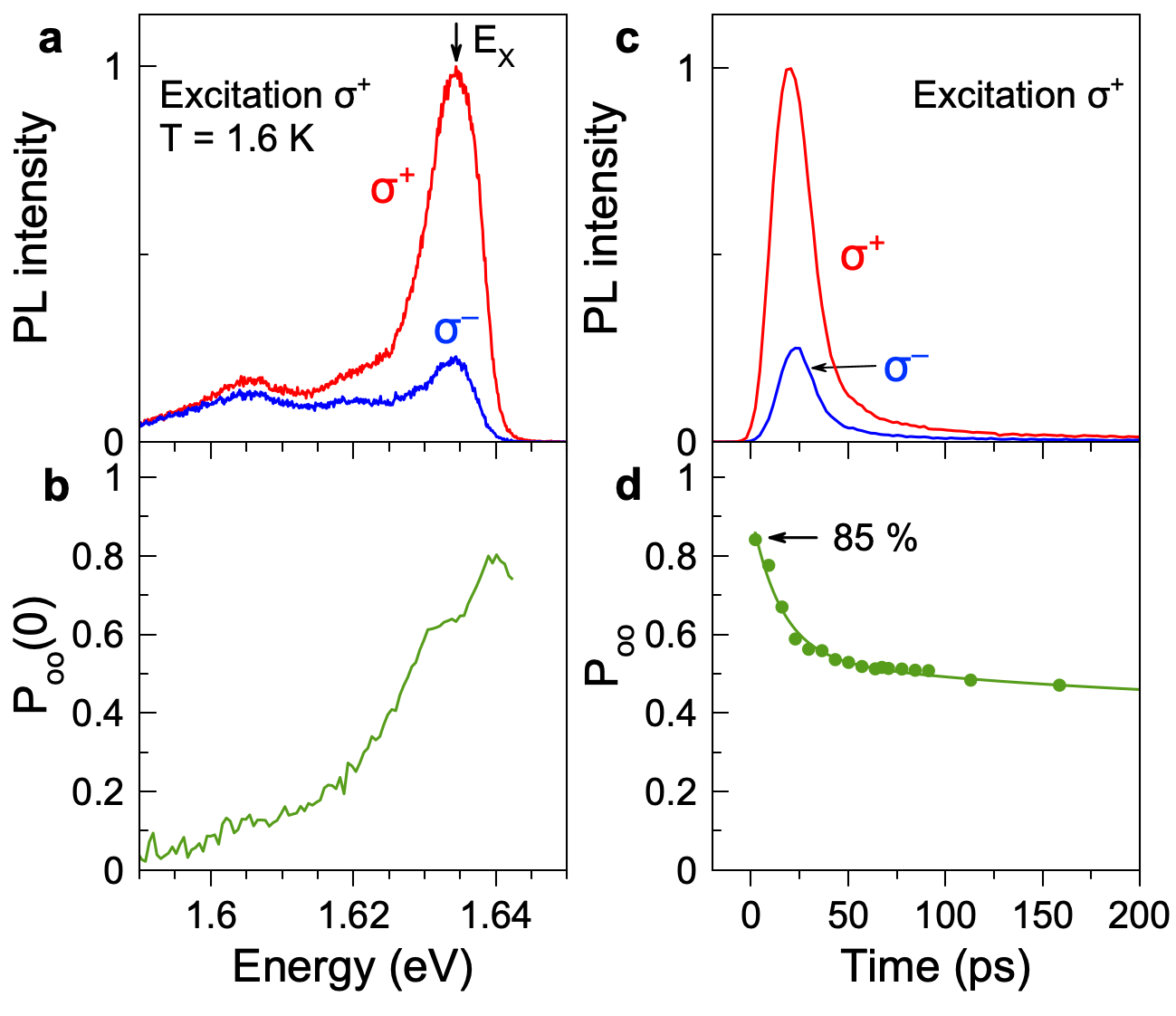}
\caption{\label{fig3} Optical orientation of excitons in MAPbI$_3$ crystal. (a) PL spectra right after $\sigma^{+}$ polarized excitation with 200~fs laser pulses ($t = 0$~ps, $E_\text{exc} = 1.698$~eV, excitation density of 10~mW/cm$^2$), measured in  $\sigma^{+}$ (red line) and $\sigma^{-}$ (blue line) polarization. $T= 1.6$~K. (b) Spectral dependence of the optical orientation degree calculated for the data in panel (a). (c) PL dynamics measured in $\sigma^{+}$ (red line) and $\sigma^{-}$ (blue line) polarization for $\sigma^{+}$ polarized excitation. The signal is detected in the spectral range of $\delta_\text{X} = 1.635-1.645$~eV. (d) Dynamics of the optical orientation degree $P_\text{oo}(t)$.}
\end{center}
\end{figure}

Figure~\ref{fig3}(a) shows the $\sigma^+$ and $\sigma^-$ circularly polarized PL spectra right after $\sigma^+$ polarized excitation pulse action ($t = 0$\,ps). At the exciton energy, the $\sigma^+$ polarized emission is considerably stronger than the $\sigma^-$ polarized one, evidencing a large degree of optical orientation. The latter we define as:
\begin{equation}
\label{eq1}
P_{\rm{oo}} = \frac{I^{++} - I^{+-}}{I^{++} + I^{+-}}\,.
\end{equation}
Here $I^{++}$ and $I^{+-}$ are the intensities of the $\sigma^+$ and $\sigma^-$ polarized emission measured for $\sigma^+$ polarized excitation. The spectral dependence of $P_{\rm{oo}}$ calculated by Eq.~\eqref{eq1} is presented in Figure~\ref{fig3}(b). The maximal optical orientation degree of 0.85 (or 85\%) is observed at the high energy flank of the exciton at 1.640~eV. Note that we reported recently the same giant optical orientation of excitons for FA$_{0.9}$Cs$_{0.1}$PbI$_{2.8}$Br$_{0.2}$ crystals~\cite{XOO2024} with almost cubic crystal symmetry at cryogenic temperatures, as confirmed by isotropic carrier $g$-factors~\cite{kirstein2022nc,kirstein2022AM}.

The dynamics of the $\sigma^+$ and $\sigma^-$ circularly polarized exciton emission after the $\sigma^+$ polarized pulse are shown in Figure~\ref{fig3}(c). The $\sigma^+$ polarized dynamics have a larger amplitude than in the opposite helicity, evidencing the large optical orientation degree. The dynamics of $P_{\rm{oo}}(t)$ are plotted in Figure~\ref{fig3}(d). The maximal optical orientation degree of 0.85 is observed right after the pulse action. The optical orientation decreases with an initial decay time of about 20~ps, saturating at about 0.45 for delays exceeding 200~ps, where excitons have already recombined. This behavior is similar to the recently reported optical orientation for FA$_{0.9}$Cs$_{0.1}$PbI$_{2.8}$Br$_{0.2}$ in Ref.~\cite{XOO2024}, where it is shown that the decrease of  $P_{\rm{oo}}$ is not related to exciton spin relaxation, but is determined by exciton recombination. In this case, two recombination processes coexist at the same energy: fast exciton recombination and slower recombination of spatially separated electrons and holes. The exciton spin relaxation time strongly exceeds its recombination time (in FA$_{0.9}$Cs$_{0.1}$PbI$_{2.8}$Br$_{0.2}$ by a factor of 5). The observed decrease of $P_{\rm{oo}}$ reflects the fact that all highly-polarized excitons have recombined, and $P_{\rm{oo}}$ at longer delays is contributed by the optical orientation of carriers with a smaller degree of $P_{\rm{oo}}$~\cite{COO2024}.   

\subsection{Robustness of optical orientation against optical detuning}

The surprising peculiarity of optical orientation in FA$_{0.9}$Cs$_{0.1}$PbI$_{2.8}$Br$_{0.2}$ crystals is its outstanding robustness against detuning of the excitation energy from the exciton resonance~\cite{XOO2024}. This gives strong experimental evidence that the spin splitting of the conduction and valence bands at finite wave vectors, which might be provided by the Rashba and Dresselhaus effects, are absent in materials with almost cubic crystal symmetry. Therefore, it evidences that the lattice has inversion symmetry, as expected for the crystal structure of cubic perovskites. This raises immediately the question whether the reduction of crystal symmetry to orthorhombic induces mechanisms that break the inversion symmetry. To answer, MAPbI$_3$ is the optimal material as its band gap is close to that of FA$_{0.9}$Cs$_{0.1}$PbI$_{2.8}$Br$_{0.2}$, since the gap is controlled by the iodine anions, but the crystal symmetry is reduced from cubic to orthorhombic below the phase transition temperature of 160$^\circ$C.  

\begin{figure*}[htb]
\begin{center}
\includegraphics[width = 15cm]{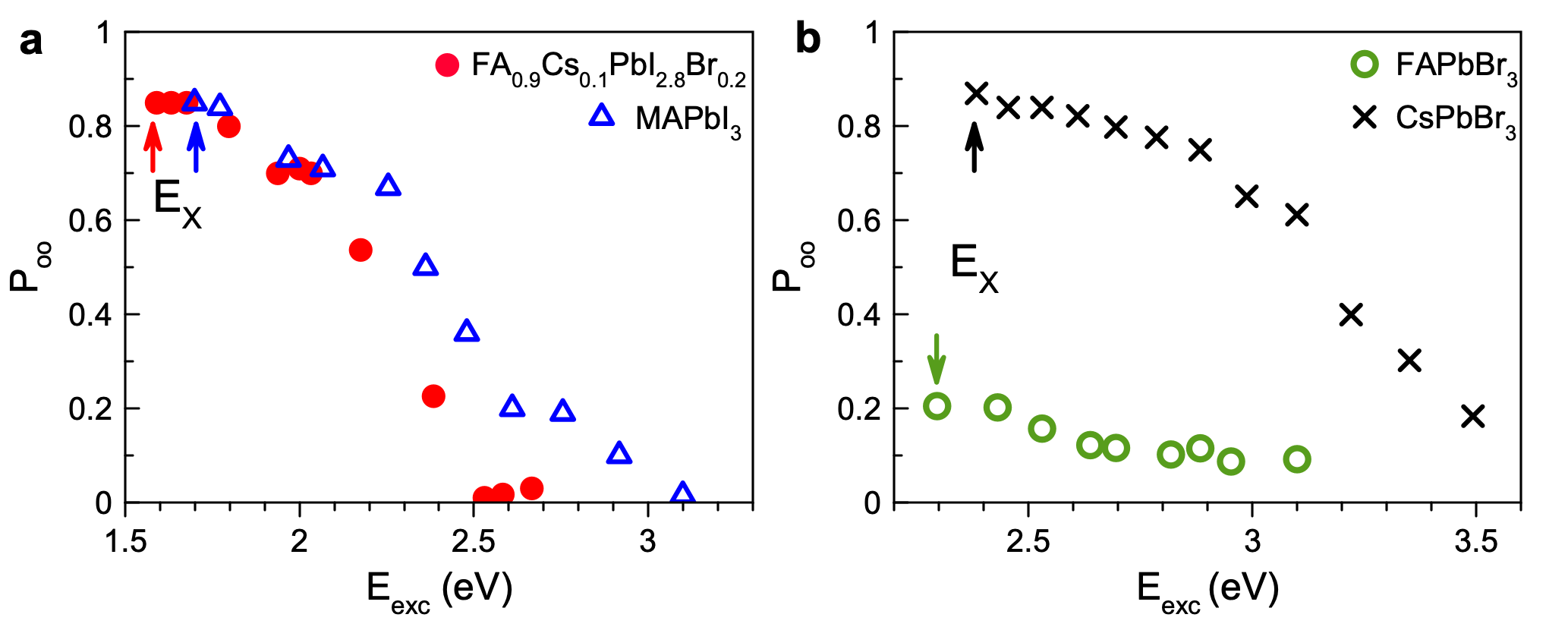}
\caption{\label{fig4} Optical orientation of excitons versus laser excitation energy. Dependence of $P_\text{oo} (t = 0)$ on $E_{\rm exc}$ for (a) MAPbI$_3$ (blue) and FA$_{0.9}$Cs$_{0.1}$PbI$_{2.8}$Br$_{0.2}$ (red) as well as (b) FAPbBr$_3$ (green) and  CsPbBr$_3$  (black). The excitation density is 10~mW/cm$^2$ at $T = 1.6$~K. Arrows mark the exciton resonances (E$_\text{X}$) of corresponding materials.
}
\end{center}
\end{figure*}

\begin{figure*}[htb]
\begin{center}
\includegraphics[width = 15cm]{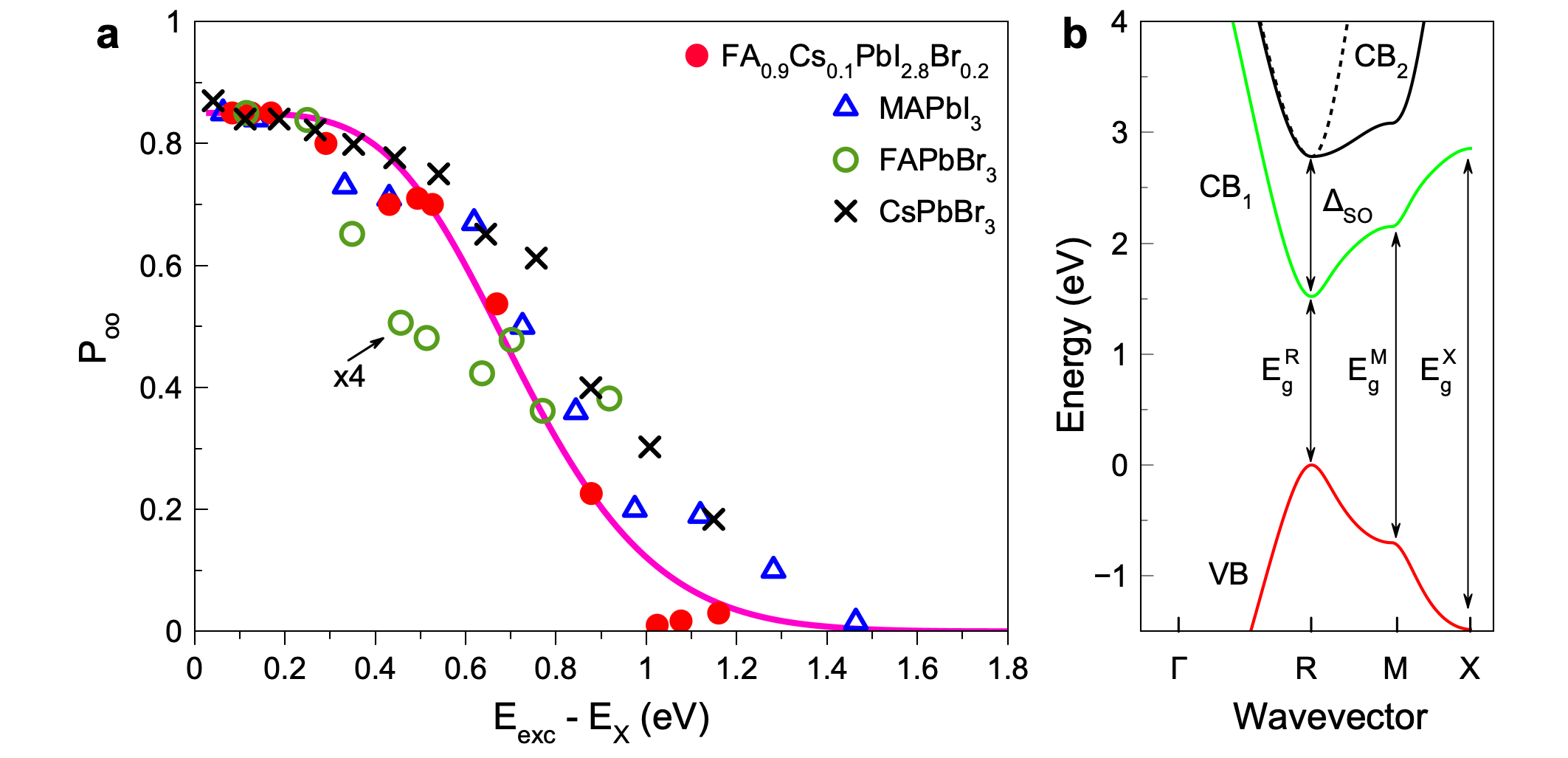}
\caption{\label{fig5} (a) Optical orientation degree of excitons in various perovskite crystals as function of the optical detuning from the exciton resonance. The data are taken from Figure~\ref{fig4} with the exciton energy set to zero. The values for FAPbBr$_3$ are multiplied by a factor of four for better comparison. The pink line is the theoretical dependence calculated with account for the Elliott-Yafet spin relaxation due to interaction with longitudinal optical phonons, see Ref.~\onlinecite{XOO2024}. (b)  Calculated band diagram of bulk FA$_{0.9}$Cs$_{0.1}$PbI$_{2.8}$Br$_{0.2}$ along the $\Gamma \to  R \to M \to  X$ path. The vertical lines indicate the transition energies at the $R$, $M$, and $X$ points from the valence band (VB) to the conduction band (CB$_1$). The light- and heavy-electron bands CB$_2$ are shifted from CB$_1$ by the spin-orbit splitting ($\Delta_\text{SO}$).}
\end{center}
\end{figure*}

In Figure~\ref{fig4}(a) the dependence of the exciton optical orientation degree on the excitation energy detuned to higher values relative to the exciton resonance is shown for the MAPbI$_3$ crystal by the blue triangles. $P_\text{oo}$ starts at 0.85 for small excitation energies relative to $E_{\rm X}$ and decreases weakly up to about 0.3~eV detuning. The decrease becomes prominent for larger detunings and $P_\text{oo}$ reaches zero for a detuning of about 1.4~eV. This behavior is similar to the findings for almost cubic FA$_{0.9}$Cs$_{0.1}$PbI$_{2.8}$Br$_{0.2}$ crystals, shown here by the red circles. As the band gaps of these materials differ by about 0.1~eV, we plot for better comparison the data in Figure~\ref{fig5}(a) as function of the detuning energy from the exciton resonance ($E_{\rm exc}-E_{\rm X}$), so that $E_{\rm X} = 0$. The dependences for the two materials coincide remarkably well with each other. From these data, we conclude that the reduction of crystal symmetry from almost cubic to orthorhombic does not result in  activation of any additional spin relaxation mechanisms. The main candidate would be the Dyakonov-Perel mechanism which should become efficient due to a spin splitting of the electron and hole states at finite wave vectors in crystals without inversion symmetry combined with a strong spin-orbit interaction. Therefore, one can conclude that the spatial inversion symmetry is maintained for the orthorhombic MAPbI$_3$ crystal.  

To collect comprehensive information on the whole class of lead halide perovskite semiconductors, we examine experimentally two more materials with bromine anions resulting in larger band gaps. These are the almost cubic FAPbBr$_{3}$ and the orthorhombic CsPbBr$_{3}$ crystals. Their optical spectra are shown in Figures~\ref{fig1}(c,d). In the first material, the electron and hole $g$-factors are almost isotropic~\cite{Kirstein_2024_FAPbBr3}, and in the second one the $g$-factors show a pronounced anisotropy~\cite{kirstein2022nc}. The optical orientation of excitons in these crystals measured at various excitation energies is shown in Figure~\ref{fig4}(b). For small detunings from $E_{\rm X}$, $P_\text{oo} (t = 0)$ reaches 0.85 in CsPbBr$_{3}$ and 0.20 in FAPbBr$_{3}$. Despite the difference in the absolute values, their dependence on excitation energy is very similar, as can be better seen in Figure~\ref{fig5}(a), where the data for FAPbBr$_{3}$ are multiplied by a factor of 4 to be normalized to 0.85 value in CsPbBr$_{3}$. The basically identical dependence on the detuning allows us to conclude that the difference in the optical orientation degree is provided by a more efficient spin relaxation of cold excitons in FAPbBr$_{3}$. We suggest that the responsible mechanism is provided by exciton interaction with the nuclear spin fluctuations~\cite{COO2024,Kirstein2023_DNSS}, 
which requires, however, further investigations beyond the scope of the present study.   

In Figure~\ref{fig5}(a) the exciton optical orientation in the four studied materials is compared. All of them show a similar dependence on the excitation energy detuning, which evidences that the dependence is determined by the band structure of the lead halide perovskites around the band gap to energies $0.5-0.8$~eV higher than the conduction and valence band extrema. Here, we keep in mind that the effective masses of electrons and holes in the lead halide perovskites are about equal to each other. Note that the investigated excitation energy detuning is  smaller than the spin-orbit-splitting $\Delta_{\rm SO}$ to the heavy- and light-electron bands of about $1.2-1.5$~eV and the energy gap at the M-point, which exceeds the band gap at the R-point by about 1.33~eV, see Figure~\ref{fig5}(b).

\subsection{Discussion}

The optical orientation degree is determined by two factors: (i) the efficiency of spin initialization induced by circularly polarized photons for the interband optical transition and (ii) the loss of spin polarization during the exciton lifetime $\tau_{\rm R}$. The latter is controlled by the factor $\tau_{\rm s}/(\tau_{\rm s}+\tau_{\rm R})$ that reduces the initial optical orientation, where $\tau_{\rm s}$ is the exciton spin relaxation time. The band structure of lead halide perovskites is favorable for obtaining high optical orientation close to the ultimate limit of 100\%~\cite{XOO2024}. In the vicinity of the band gap, i.e., in the detuning range up to about 0.3~eV, the spins can be optically oriented with 100\% efficiency. Also, due to the presence of inversion symmetry of the crystal lattice, spin splittings of the bands through the spin-orbit interaction are absent, suppressing the spin relaxation by the Dyakonov-Perel mechanism. Instead, the spin relaxation is governed by the rather inefficient Elliott-Yafet mechanism and the electron-hole exchange interaction in the exciton. These expectations are in agreement with the observed high optical orientation degree of 0.85 measured for three of the studied materials. We can safely conclude that  $\tau_{\rm s} \gg \tau_{\rm R}$ for the excitons in MAPbI$_3$, FA$_{0.9}$Cs$_{0.1}$PbI$_{2.8}$Br$_{0.2}$, and CsPbBr$_3$.

An in-depth theoretical analysis based on tight-binding calculations of the spin initialization efficiency at large detunings has been performed in Ref.~\onlinecite{XOO2024}. The line in Figure~\ref{fig5}(a) shows the calculated dependence, which matches well the experimental data for all four materials. The analysis shows that the decrease of optical orientation at larger detunings originates from the deviation of the optical helicity selection rules from the strict ones at the R-point. Also, additional depolarization comes from the spin-flip scattering of excitons during the energy relaxation provided by the Elliott-Yafet mechanism. 

It is well established that CsPbBr$_3$ at low temperatures has an orthorhombic crystal structure with space group \#62. This group has an inversion center, so that no spin splitting in the bands is expected. However, samples with organic cations (MA$^+$ and FA$^+$) may show a lower local symmetry due to the oriented organic cations~\cite{Kepenekian2017}. Close similarities of spin properties found for Cs, MA, and FA-based materials let us conclude that the latter effect does not contribute to spin dynamics.

We emphasize that the experimental technique used in our study delivers information about the exciton properties inside a crystal and not at its surface, where excitons are typically destroyed. Therefore, our experiments cannot detect a possible breaking of the inversion symmetry within the few monolayers close to the crystal surface, as reported recently, e.g., for a MAPbBr$_3$ single crystal by the circular photogalvanic effect~\cite{Huang2021}.

\section{Conclusion}

We have demonstrated that the exciton spins can be optically oriented in FA$_{0.9}$Cs$_{0.1}$PbI$_{2.8}$Br$_{0.2}$, MAPbI$_3$, CsPbBr$_3$, and FAPbBr$_3$ crystals and in most of them a very high degree of optical orientation of 85\% is reached at cryogenic temperatures. For different perovskite compositions and symmetries, the optical orientation is robust against detuning of the laser excitation energy from the exciton resonance. We find that the exciton spin relaxation mechanisms are not influenced by the symmetry reduction from almost cubic to orthorhombic, evidencing that the spatial inversion symmetry inherent to cubic perovskite crystals is not broken for orthorhombic crystals.

In bulk perovskite crystals, the inversion symmetry may be absent at their surface, and in polycrystalline films, it may get lost at the crystalline boundaries. Also, external factors, like axial strain and electric field, may break the inversion symmetry. The surface effect may become detrimental for perovskite nanocrystals and two-dimensional materials, where a Rashba spin splitting has been reported, based on experimental data~\cite{Lafalce2022,zhai2017,becker2018} and theoretical predictions~\cite{becker2018,sercel2023}. We are convinced that the optical orientation technique is an extremely useful tool for examining the effects associated with the crystal symmetry in perovskite nanostructures.

\section{Experimental Section}

\textbf{Samples:}  We study single lead halide perovskite crystals with different band gaps. The crystals are grown out of solution using the inverse temperature crystallization (ITC) technique~\cite{Dirin2016,nazarenko2017,hocker2021}. For the FA$_{0.9}$Cs$_{0.1}$PbI$_{2.8}$Br$_{0.2}$, FAPbBr$_3$, and CsPbBr$_3$ crystals studied here, details of the synthesis are given in Ref.~\onlinecite{kirstein2022nc}. The MAPbI$_3$ crystal was synthesized from PbI$_2$ and MAI perovskite precursors between two polytetrafluoroethylene coated glasses, following Ref.~\cite{Yang2022}. The sample has a square shape of $1\times1$~mm$^2$ area in the (001) crystallographic plane with a thickness of 30~$\mu$m. 

The hybrid organic-inorganic compounds FA$_{0.9}$Cs$_{0.1}$PbI$_{2.8}$Br$_{0.2}$ ($E_\text{g} = 1.520$\,eV) and MAPbI$_{3}$ (1.652~eV) have band gap energies close to the near-infrared at $T = 1.6$~K temperature. Replacing the iodine halogen with bromine results in a blue shift of the band gap for FAPbBr$_{3}$ (2.216~eV). To develop a complete picture, we also study the fully inorganic perovskite CsPbBr$_{3}$ (2.359~eV).  The $g$-factor tensor of electrons and holes measured at the cryogenic temperatures is isotropic for FA$_{0.9}$Cs$_{0.1}$PbI$_{2.8}$Br$_{0.2}$ and FAPbBr$_{3}$, reflecting their almost cubic crystal symmetry~\cite{kirstein2022nc}. Change of the cation from FA$^+$ to MA$^+$ and Cs$^+$ leads to the symmetry reduction, evidenced by a strong $g$-factor anisotropy of carriers~\cite{kirstein2022nc}.

\textbf{Optical measurements:} 
For the low-temperature optical measurements, we use a liquid helium cryostat with the temperature variable from 1.6\,K up to 300\,K. At $T=1.6$\,K, the sample is immersed in superfluid helium. 

\textbf{Photoluminescence and reflectivity measurements:} 
The time-integrated photoluminescence (PL) and reflectivity spectra are measured with a 0.5\,m spectrometer equipped with a charge-coupled-device (CCD) camera. We use a halogen lamp as light source for the reflectivity measurements.

\textbf{Time-resolved photoluminescence:}
The spectrally-resolved PL dynamics are measured using a spectrometer with a 300 grooves/mm diffraction grating and a streak camera with 10~ps time resolution. Laser pulses with 200\,fs duration and central photon energies from 1.7\,eV (730\,nm) to 3.5\,eV (350\,nm) taken from a tunable Coherent Chameleon Discovery laser (repetition rate of 80~MHz) with a second harmonic generation unit are used for PL excitation. To study the effect of optical orientation, we use circular polarized ($\sigma^+/\sigma^-$) excitation light and subsequently analyze the circular polarization of the emission.

\subsection*{Data Availability Statement}
The data presented in this paper are available from the corresponding authors upon reasonable request.

\subsection*{Acknowledgements}
The authors are thankful to Al. L. Efros and M. M. Glazov for fruitful discussions. We acknowledge the financial support by the Deutsche Forschungsgemeinschaft via the SPP2196 Priority Program (projects YA65/28-1, no. 527080192, and AK40/13-1, no. 506623857). I.V.K. acknowledges the support of the Deutsche Forschungsgemeinschaft (project KA 6253/1-1, no. 534406322). The work at ETH Z\"urich (B.T., O.H., D.N.D., and M.V.K.) was financially supported by the Swiss National Science Foundation (grant agreement 200020E 217589) through the DFG-SNSF bilateral program and by ETH Z\"urich through ETH+ Project SynMatLab.

\section*{Author information}
 \subsection*{Corresponding Authors}
 \textbf{Nataliia E. Kopteva} -- Experimentelle Physik 2, Technische Universit\"at Dortmund, 44227 Dortmund, Germany; orcid.org/0000-0003-0865-0393; Email: natalia.kopteva@tu-dortmund.de\\
 
 \textbf{Dmitri R. Yakovlev} -- Experimentelle Physik 2, Technische Universit\"at Dortmund, 44227 Dortmund, Germany; orcid.org/0000-0001-7349-2745; Email: dmitri.yakovlev@tu-dortmund.de\\

\subsection*{Authors}
 \textbf{Ey\"up Yalcin} -- Experimentelle Physik 2, Technische Universit\"at Dortmund, 44227 Dortmund, Germany; orcid.org/0000-0003-2891-4173\\
 
 \textbf{Ina V. Kalitukha} -- Experimentelle Physik 2, Technische Universit\"at Dortmund, 44227 Dortmund, Germany; orcid.org/0000-0003-2153-6667\\
 
 \textbf{Ilya A. Akimov} -- Experimentelle Physik 2, Technische Universit\"at Dortmund, 44227 Dortmund, Germany; orcid.org/0000-0002-2035-2324\\
 
  \textbf{Mikhail O. Nestoklon} -- Experimentelle Physik 2, Technische Universit\"at Dortmund, 44227 Dortmund, Germany; orcid.org/0000-0002-0454-342X\\
 
 \textbf{Bekir T\"uredi} -- Laboratory of Inorganic Chemistry, Department of Chemistry and Applied Biosciences,  ETH Z\"{u}rich, CH-8093 Z\"{u}rich, Switzerland; Laboratory for Thin Films and Photovoltaics, Empa-Swiss Federal Laboratories for Materials Science and Technology, CH-8600 D\"{u}bendorf, Switzerland; orcid.org/0000-0003-2208-0737\\
 
 \textbf{Oleh Hordiichuk} -- Laboratory of Inorganic Chemistry, Department of Chemistry and Applied Biosciences,  ETH Z\"{u}rich, CH-8093 Z\"{u}rich, Switzerland; Laboratory for Thin Films and Photovoltaics, Empa-Swiss Federal Laboratories for Materials Science and Technology, CH-8600 D\"{u}bendorf, Switzerland; orcid.org/0000-0001-7679-4423\\
 
 \textbf{Dmitry N. Dirin} -- Laboratory of Inorganic Chemistry, Department of Chemistry and Applied Biosciences,  ETH Z\"{u}rich, CH-8093 Z\"{u}rich, Switzerland; Laboratory for Thin Films and Photovoltaics, Empa-Swiss Federal Laboratories for Materials Science and Technology, CH-8600 D\"{u}bendorf, Switzerland; orcid.org/0000-0002-5187-4555\\
 
 \textbf{Maksym~V.~Kovalenko} -- Laboratory of Inorganic Chemistry, Department of Chemistry and Applied Biosciences,  ETH Z\"{u}rich, CH-8093 Z\"{u}rich, Switzerland; Laboratory for Thin Films and Photovoltaics, Empa-Swiss Federal Laboratories for Materials Science and Technology, CH-8600 D\"{u}bendorf, Switzerland; orcid.org/0000-0002-6396-8938\\
 
 \textbf{Manfred Bayer} -- Experimentelle Physik 2, Technische Universit\"at Dortmund, 44227 Dortmund, Germany; orcid.org/0000-0002-0893-5949\\



\section*{References}

\begin{thebibliography}{99}

\bibitem{jena2019} A.~K. Jena, A. Kulkarni, T. Miyasaka, 
Halide perovskite photovoltaics: Background, status, and future prospects.
\textit{Chem. Rev.} \textbf{2019}, \textit{5}, {3036--3103}.

\bibitem{Vinattieri2021_book} \textit{Halide Perovskites for Photonics}, eds.  A. Vinattieri and G. Giorgi,  (AIP Publishing, Melville, New York, \textbf{2021}).

\bibitem{Vardeny2022_book} \textit{Hybrid Organic Inorganic Perovskites: Physical Properties and Applications}, eds. Z. V. Vardeny  and  M. C. Beard, (World Scientific, \textbf{2022}).

\bibitem{simanes2024} M. Simenas, A. Gagor, J. Banys, M. Maczka, Phase transitions and dynamics in mixed three- and low-
dimensional lead halide perovskites
\textit{Chem. Rev.} \textbf{2024},  \textit{124}, 2281.

\bibitem{wang2019} J. Wang, C. Zhang, H. Liu,  R. McLaughlin, Y. Zhai, S. R. Vardeny, X. Liu, S.  McGill,  D. Semenov, H. Guo, R. Tsuchikawa, V. V. Deshpande, D. Sun,  Z. V.  Vardeny, 
Spin-optoelectronic devices based on hybrid organic-inorganic trihalide perovskites.
\textit{Nat. Commun.} \textbf{2019},  \textit{10}, 129.

\bibitem{ning2020} W. Ning,  J. Bao, Y. Puttisong, F. Moro, L. Kobera, S. Shimono, L. Wang, F. Ji, M. Cuartero, S. Kawaguchi, S. Abbrent, H. Ishibashi, R. De Marco, I. A. Bouianova, G. A. Crespo, Y. Kubota, J. Brus, D. Y. Chung, L. Sun, W. M. Chen, M. G. Kanatzidis, F. Gao, 
Magnetizing lead free halide double perovskites.
\textit{Science Advances}  \textbf{2020}, \textit{6}, {eabb5381}.

\bibitem{kim2021} Y.-H. Kim, Y. Zhai, H. Lu, X. Pan, C. Xiao, E. A. Gaulding, S. P. Harvey, J. J. Berry,  Z.~V. Vardeny, J. M. Luther,  M. C. Beard, 
Chiral-induced spin selectivity enables a room-temperature spin light-emitting diode.
\textit{Science}  \textbf{2021},	\textit{371}, {1129}.

\bibitem{XOO2024} N. E. Kopteva, D. R. Yakovlev, E. Yalcin, I. A. Akimov, M. O. Nestoklon, M. M. Glazov, M. Kotur, D. Kudlacik, E. A. Zhukov, E. Kirstein, O. Hordiichuk, D. N. Dirin, M. V. Kovalenko, M. Bayer, 
Highly-polarized emission provided by giant optical orientation of exciton spins in lead halide perovskite crystals. \textit{Adv. Sci.} \textbf{2024}, \textit{11}, 2403691.

\bibitem{COO2024} D.~Kudlacik, N.~E.~Kopteva, M.~Kotur, D.~R.~Yakovlev, K.~V.~Kavokin, C.~Harkort, M.~Karzel, E.~A.~Zhukov, E.~Evers, V.~V.~Belykh, M.~Bayer, Optical spin orientation of localized electrons and holes interacting with nuclei in a FA$_{0.9}$Cs$_{0.1}$PbI$_{2.8}$Br$_{0.2}$ perovskite crystal. \textit{ACS Photonics} \textbf{2024}, \textit{11}, 2757.

\bibitem{Giovanni2015} D. Giovanni, H. Ma, J. Chua, M. Gr\"atzel, R. Ramesh, S. Mhaisalkar, N. Mathews, T. Ch.  Sum, 
Highly spin-polarized carrier dynamics and ultralarge photoinduced magnetization in {CH}$_3${NH}$_3${PbI}$_3$ perovskite thin films.
\textit{Nano Lett.} \textbf{2015}, \textit{15}, {1553-1558}.

\bibitem{belykh2019} V. V. Belykh, D. R. Yakovlev, M. M. Glazov, P. S. Grigoryev, M. Hussain, J. Rautert,  D. N. Dirin, M. V.  Kovalenko,  M. Bayer,
Coherent spin dynamics of electrons and holes in CsPbBr$_3$ perovskite crystals.
\textit{Nat. Commun.} \textbf{2019},	\textit{10}, {673}.

\bibitem{odenthal2017} P. Odenthal, W. Talmadge, N. Gundlach, R. Wang, C. Zhang, D. Sun, Z.-G. Yu, V. Z. Vardeny,  Y. S. Li,  
Spin-polarized exciton quantum beating in hybrid organic--inorganic perovskites. 
\textit{Nat. Phys.} \textbf{2017}, \textit{13}, {894}.

\bibitem{kirstein2022nc} E. Kirstein, D. R. Yakovlev, M. M. Glazov, E. A. Zhukov, D. Kudlacik, I. V. Kalitukha, V. F. Sapega, G. S. Dimitriev, M. A. Semina, M. O. Nestoklon, E. L. Ivchenko, N. E. Kopteva,
D. N. Dirin, O. Nazarenko, M. V. Kovalenko, A. Baumann, J. H\"ocker, V. Dyakonov, M. Bayer, 
The Land\'e factors of electrons and holes in lead halide perovskites: universal dependence on the band gap.
\textit{Nat. Commun.} \textbf{2022}, \textit{13}, {3062}.

\bibitem{kirstein2022AM} E. Kirstein, D. R. Yakovlev, M. M. Glazov, E. Evers, E. A. Zhukov,
V. V. Belykh, N. E. Kopteva, D. Kudlacik, O. Nazarenko, D. N. Dirin, M. V. Kovalenko, M. Bayer,
Lead-dominated hyperfine interaction impacting the carrier spin dynamics in halide perovskites.
\textit{Advanced Materials} \textbf{2022},  \textit{34}, 2105263.

\bibitem{Belykh2022}  V. V. Belykh, M. L. Skorikov, E. V. Kulebyakina, E. V. Kolobkova, M. S. Kuznetsova, M. M. Glazov,  D. R. Yakovlev,
Submillisecond spin relaxation in CsPb(Cl,Br)$_3$ perovskite nanocrystals in a glass matrix.  
\textit{Nano Lett.} \textbf{2022}, \textit{22}, 4583--4588.

\bibitem{Kirstein2023_DNSS} E. Kirstein, D. S. Smirnov, E. A. Zhukov, D. R. Yakovlev, N. E. Kopteva, D. N. Dirin, O. Hordiichuk, M. V. Kovalenko,  M. Bayer, 
The squeezed dark nuclear spin state in lead halide perovskites, 
\textit{Nat. Commun.}  \textbf{2023},  \textit{14}, 6683.

\bibitem{Manchon2015} A. Manchon, H. C. Koo, J. Nitta, S. M. Frolov, R. A. Duine, 
New perspectives for Rashba spin–orbit coupling.
\textit{Nat. Mater.} \textbf{2015}, \textit{14}, 871.

\bibitem{Kepenekian2015} M. Kepenekian, R. Robles, C. Katan, D. Sapori, L. Pedesseau, J.  Even, 
{R}ashba and {D}resselhaus effects in hybrid organic-inorganic perovskites: from basics to devices.
\textit{ACS Nano} \textbf{2015}, \textit{9}, 11557.

\bibitem{Kepenekian2017} M. Kepenekian, J.  Even, 
{R}ashba and {D}resselhaus couplings in halide perovskites: accomplishments and opportunities for spintronics and spin-orbitronics.
\textit{J. Phys. Chem. Lett.} \textbf{2017}, \textit{8}, 3362.

\bibitem{Mosconi2017} E. Mosconi, T. Etienne, F. De Angelis, 
Rashba band splitting in organohalide lead perovskites: Bulk and surface effects.
\textit{J. Phys. Chem. Lett.} \textbf{2017}, \textit{8}, 2247.

\bibitem{Leppert2016} L. Leppert, S. E. Reyes-Lillo, J. B. Neaton, 
Electric field- and strain-induced Rashba effect in hybrid halide perovskites.
\textit{J. Phys. Chem. Lett.} \textbf{2016}, \textit{7}, 3683.

\bibitem{Kim2014} M. Kim, J. Im, A. J. Freeman, J. Ihm, H. Jin, 
Switchable $S = 1/2$ and $J = 1/2$ Rashba bands in ferroelectric halide perovskites.
\textit{PNAS} \textbf{2014}, \textit{111}, 6900.

\bibitem{Yu2019} Zh.-G. Yu, Y. S. Li,
Unraveling the spin relaxation mechanism in hybrid organic–inorganic perovskites.
\textit{J. Phys. Chem. C} \textbf{2019}, \textit{123}, 14701.

\bibitem{Marronnier2018} A. Marronnier, G. Roma, M. A. Carignano, Y. Bonnassieux, C. Katan, J. Even, E. Mosconi, F. De Angelis,
Influence of disorder and anharmonic fluctuations on the dynamical Rashba effect in purely inorganic lead-halide perovskites.
\textit{J. Phys. Chem. C} \textbf{2019}, \textit{123}, 291.

\bibitem{McKechnie2018} S. McKechnie, J. M. Frost, D. Pashov, P. Azarhoosh, A. Walsh, M. van Schilfgaarde, 
Dynamic symmetry breaking and spin splitting in metal halide perovskites.
\textit{Phys. Rev. B} \textbf{2018}, \textit{98}, 085108.

\bibitem{Niesner2016} D. Niesner, M. Wilhelm, I. Levchuk, A. Osvet, S. Shrestha, M. Batentschuk, C. Brabec,  T. Fauster, 
Giant Rashba splitting in CH$_3$NH$_3$PbBr$_3$ organic-inorganic perovskite.
\textit{Phys. Rev. Lett.} \textbf{2016}, \textit{117}, 126401.

\bibitem{Lafalce2022} E. Lafalce, E. Amerling, Z.-G. Yu, P. C. Sercel, L. Whittaker-Brooks, Z. V. Vardeny, 
Rashba splitting in organic–inorganic lead–halide perovskites revealed through two-photon absorption spectroscopy.
\textit{Nat. Commun.} \textbf{2022}, \textit{13}, 483.

\bibitem{Huang2021} Z. Huang, Sh. R. Vardeny, T. Wang, Z. Ahmad, A. Chanana, E. Vetter, Sh. Yang, X. Liu,  G. Galli, A. Amassian,  Z. V. Vardeny, D.  Sun, 
Observation of spatially resolved Rashba states on the surface of CH$_3$NH$_3$PbBr$_3$ single crystals. 
\textit{Appl. Phys. Rev.} \textbf{2021},  \textit{8}, 031408.

\bibitem{Sajedi2020} M. Sajedi, M. Krivenkov, D. Marchenko,  A. Varykhalov, J. S\'anchez-Barriga, E. D. L. Rienks, O. Rader,  
 Absence of a giant {R}ashba effect in the valence band of lead halide perovskites.
\textit{Phys. Rev. B} \textbf{2020}, \textit{102}, 081116.

\bibitem{Frohna2018} K. Frohna,  T. Deshpande, J. Harter, W. Peng, B. A. Barker, J. B. Neaton, S. G. Louie,  O. M. Bark, D. Hsieh, M. Bernardi,  
Inversion symmetry and bulk Rashba effect in methylammonium lead iodide perovskite single crystals.
\textit{Nat. Commun.} \textbf{2018}, \textit{9}, 1829.

\bibitem{Niesner2018} D. Niesner, M. Hauck, Sh. Shrestha, I. Levchukc,  G. J. Matt, A. Osvet, M. Batentschuk,  Ch. Brabec,  H. B. Weber, Th.  Fauster, 
Structural ﬂuctuations cause spin-split states in tetragonal (CH$_3$NH$_3$)PbI$_3$ as evidenced by the circular photogalvanic effect.
\textit{PNAS} \textbf{2018}, \textit{115}, 9509.

\bibitem{herz2017} A. D. Wright, R. L. Milot, G. E. Eperon, H. J. Snaith, M.B Johnston, L.M. Herz,  Band-tail recombination in hybrid lead iodide perovskite. \textit{Adv. Funct. Mater.} \textbf{2017}, \textit{27}, {1700860}.

\bibitem{deQuilettes2019_si} D.~W. deQuilettes, K. Frohna, D. Emin, T. Kirchartz, V. Bulovic, D.~S. Ginger, S.~D. Stranks, 
Charge-carrier recombination in halide perovskites.
\textit{Chem. Rev.} \textbf{2019}, \textit{119}, 11007–-11019.

\bibitem{Yakovlev2024} D. R. Yakovlev, S. A. Crooker, M. A. Semina, J. Rautert, J. Mund, D. N. Dirin, M. V. Kovalenko,  M. Bayer, 
Exciton–polaritons in {C}s{P}b{B}r$_3$ crystals revealed by optical reflectivity in high magnetic fields and two-photon spectroscopy.
\textit{Phys. Stat. Sol. RRL} \textbf{2024}, \textit{18}, 2300407.

\bibitem{kirstein2022mapi} E. Kirstein, D. R. Yakovlev, E. A. Zhukov, J. H\"ocker, V. Dyakonov, M. Bayer, 
Spin dynamics of electrons and holes interacting with nuclei in MAPbI$_3$ perovskite single crystals. 
\textit{ACS Photonics} \textbf{2022}, \textit{9}, {1375}.

\bibitem{Kirstein_2024_FAPbBr3} E. Kirstein, E. A. Zhukov, D. R. Yakovlev, N. E. Kopteva, E. Yalcin, I. A. Akimov, O. Hordiichuk, D. N. Dirin, M. V. Kovalenko,  M. Bayer,
Coherent carrier spin dynamics in FAPbBr$_3$ perovskite crystals.  
\textit{J. Phys. Chem. Lett.} \textbf{2024},  \textit{15}, 2893-2903.

\bibitem{zhai2017} Y. Zhai, S. Baniya, Ch. Zhang, J. Li, P. Haney, Ch.-X. Sheng, E. Ehrenfreund, Z. V. Vardeny,
Giant Rashba splitting in 2D organic-inorganic halide perovskites measured by transient spectroscopies.
\textit{Science Advances} \textbf{2017},  \textit{3}, e1700704.

\bibitem{becker2018} M. A. Becker, R. Vaxenburg, G. Nedelcu, P. C. Sercel, A. Shabaev, M. J. Mehl,  J. G. Michopoulos, S. G. Lambrakos, N. Bernstein, J. L. Lyons, Th. St\"oferle, R. F. Mahrt, M. V. Kovalenko,  D. J. Norris, G. Rain\'o, Al. L. Efros, 
Bright triplet excitons in caesium lead halide perovskites.
\textit{Nature} \textbf{2018}, \textit{553}, 189-193.

\bibitem{sercel2023} P. C. Sercel,  Al. L. Efros,  
Unique signatures of the Rashba effect in the magneto-optical properties of two-dimensional semiconductors. \textit{Phys. Rev. B}  \textbf{2023}, \textit{107}, 195436.


\bibitem{Dirin2016} D.~N. Dirin, I. Cherniukh, S. Yakunin, Y. Shynkarenko, M.~V. Kovalenko, 
Solution-grown {CsPbBr}$_3$ perovskite single crystals for photon detection.
\textit{Chemistry of Materials} \textbf{2016}, \textit{28}, 8470--8474.

\bibitem{hocker2021} J.~H\"ocker, F.~Brust, M.~Armer, V.~Dyakonov,
A temperature-reduced method for the rapid growth of hybrid perovskite single crystals with primary alcohols.
\textit{Cryst. Eng. Comm.} \textbf{2021}, \textit{23}, 2202--2207.

\bibitem{nazarenko2017} O.~Nazarenko, S.~Yakunin, V.~Morad, I.~Cherniukh, M.~V.~Kovalenko, 
Single crystals of cesium formamidinium lead halide perovskites: solution growth and gamma dosimetry.
\textit{NPG Asia Mater.} \textbf{2017}, \textit{9}, e373.

\bibitem{Yang2022} Ch. Yang, J. Yin, H. Li, Kh. Almasabi,  L. Guti\'errez-Arzaluz, I. Gereige, J.-L. Br\'edas,   O. M. Bakr,   O. F. Mohammed,
Engineering surface orientations for efficient and stable hybrid perovskite single-crystal solar cells.
\textit{ACS Energy Lett.}  \textbf{2022}, \textit{7}, 1544-1552.

\end{thebibliography}
\end{document}